\documentclass[preprint,12pt]{elsarticle}

\usepackage{amssymb}
 \usepackage{amsthm}

 \usepackage{lineno}

\journal{Neurocomputing}

\begin{document}

\begin{frontmatter}

\title{Phase Entrainment by Periodic Stimuli \textit{In Silico}: A Quantitative Study}


\author[inst1]{Swapna Sasi}

\affiliation[inst1]{organization={Department of Computer Science and Information Systems},
            addressline={ Birla Institute of Technology and Science, Pilani - Goa Campus}, 
            city={Zuarinagar},
            postcode={403726}, 
            state={Goa},
            country={India}}

\author[inst1]{Basabdatta Sen Bhattacharya\corref{mycorrespondingauthor}}
\cortext[mycorrespondingauthor]{Corresponding author}
\ead{basabdattab@goa.bits-pilani.ac.in}

\begin{abstract}
We present a quantitative study of phase entrainment by periodic visual stimuli in a biologically inspired neural network. The objective is to understand the neuronal population dynamics that underlie phase entrainment of brain oscillations by external stimuli, which is used for therapeutic treatment in neurological disorders, for example in Parkinsonian tremor. Yet, the neuronal dynamics underpinning such entrainment is not fully understood. Rhythmic sensory stimulation is one way of studying phase synchronisation in the brain. A recent experimental study has reported phase entrainment of brain oscillations during steady state visually evoked potentials (SSVEP), which are scalp electroencephalogram corresponding to periodic stimuli. We have simulated SSVEP-like signals corresponding to periodic pulse input to our \textit{in silico} model. We have used phase locking values, normalised Shannon entropy and conditional probability as synchronisation indices to show phase synchrony in the neuronal populations. Our experiment demonstrates that the phase synchronisation disappears with jitter in the input inter-pulse intervals, and this would not be the case if the output signal were to be the superposition of the responses to the different input signals. Thus, the phase synchronisation implies entrainment of the network response by the periodic input. Overall, our study shows the plausibility of using biologically inspired \textit{in silico} models, validated by experimental works, to understand and make testable predictions on brain entrainment as a therapeutic treatment in specific neurological disorders.
\end{abstract}




\end{frontmatter}


\section{Introduction}
\label{sec:1}
Synchrony among neuronal populations in the brain is essential for perceptual information processing as well as for building memory~\cite{Michel2020}. At the same time, unwanted synchronisation may lead to neurological conditions such as epileptic seizure~\cite{wang20}. There are different ways to measure neuronal synchrony. Our focus in this work is on phase synchronisation that is observed in steady state visually evoked potentials (SSVEP)~---~electroencephalogram (EEG) signals picked up by the scalp electrodes over the occipital and parietal lobes when a subject looks at a flickering light source. SSVEP signals have stable spectrum and high signal-to-noise ratio, which makes them popular in clinical neuroscience for studying neurological conditions such as schizophrenia, migraine and epilepsy~\cite{vialatte2010}. Yet, the neuronal dynamics underlying SSVEP are still a matter of research; the overall understanding is that it appears due to nonlinear brain dynamics~\cite{norcia2015steady}. A recent experimental study by Notbohm \emph{et al}~\cite{notbohm2016modification} addresses this issue by testing human participants with two kinds of flickering visual stimuli viz.\ pulse stimuli with periodic and jittered inter-pulse intervals. When subjected to periodic pulse stimuli, first, the fundamental frequencies in the recorded SSVEP were same as that of the stimuli; second, phase synchronisation between SSVEP and stimuli was particularly high around each subject's individual alpha frequency (IAF). (Exact peak frequency within the alpha rhythm band vary between individuals. Alpha rhythms are oscillations within the 8 -- 13 Hz frequency band that is strongest over the occipital lobe in the EEG of healthy adults in an awake and resting state with eyes closed.). However, With jittered pulse stimuli, the phase synchrony disappeared and the EEG of the participants did not reflect the input frequecy. Based on these observations, the authors conclude that SSVEP signals are a result of entrainment of brain oscillations by the periodic visual stimuli, as opposed to the superposition of oscillatory responses corresponding to each visual stimulus. Phase synchrony was measured using the Arnold tongue, normalised Shannon entropy and intermittency of phase locking. Furthermore, phase synchrony measures were between the recorded SSVEP and the visual stimuli, i.e. all observations and recordings were noninvasive. Thus, it is unclear as to what are the attributes or dynamics in specific neuron populations that play a role in phase synchronisation during SSVEP. It is known from physiological studies that the interplay between the thalamic and the cortical neuron populations are vital in generating resting state alpha rhythms~\cite{basar97,niedermeyer97}. Moreover, visual stimuli is communicated by the retina to the visual thalamus; the latter relays the information to the visual cortex (located in the occipital lobe of the brain) from where both SSVEP and resting state alpha rhythms are recorded~\cite{vialatte2010}.
It is however unknown as to whether the thalamocortical circuit dynamics of the retino-thalamo-cortical pathway can affect the generation and sustenance of SSVEP. In this regard, \textit{in silico} models of the brain, which are biologically inspired neural networks, are used to simulate and understand neuron population dynamics in the subcortical and cortical tissues. Particularly, thalamocortical \emph{in silico} models informed by the brain visual pathway are used to simulate alpha rhythms, both in normal and disease conditions~\cite{piotr04,david03}. In a recent research, an \textit{in silico} model that was originally proposed to simulate alpha rhythms, was validated by experimental observations in human participants towards understanding the nonlinear origins of SSVEP signals~\cite{Labecki2016}. 

Here, we present an \textit{in silico} study to understand the thalamocortical dynamics in the visual pathway that underpin phase synchronisation and entrainment corresponding to SSVEP as demonstrated by Notbohm \emph{et al}~\cite{notbohm2016modification}. Our goal in this work is twofold: first, to validate an \textit{in silico}  model with the experimental observations in~\cite{notbohm2016modification}; second, to use this validated \textit{in silico} model for studying the thalamocortical dynamics in the visual pathway that may play a role in the generation of phase synchronisation during SSVEP. Readers may note that by using the term `validation', we are in no way implying a direct equivalence of our model output with scalp EEG recordings. Rather, the model response is representative of physiological recordings from cortical tissue \textit{in vivo} that underlie the occipital lobe. Thus, in validating the model response with SSVEP characteristics, we conjecture that SSVEP are an attenuated representation of similar attributes in the cortical and subcortical tissues that comprise the visual pathway.

We have previously used a neural mass (\textit{in silico}) model of the thalamocortical pathway to understand alpha rhythm bio-markers in Alzheimer disease~\cite{Bhattacharya2011}. Subsequently, we have used an adapted version of the traditional neural mass model by introducing kinetic equations of neurotransmission and reception~\cite{frontiers2013}. We have been using this version of the neural mass model to understand the population dynamics underlying alpha rhythms in the visual thalamus, known as the Lateral Geniculate Nucleus (LGN) of the thalamic structure. We have shown, albeit briefly, the phase locking behaviour in the LGN output time series corresponding to periodic flickering stimuli as observed in human SSVEP~\cite{bhattacharya2016causal}. However, to understand fully the phase synchronisation in the model and the parameters that affect the associated oscillatory dynamics, quantitative measures are desirable. Towards this, we have recently collated a set of objective metrics~\cite{mahajan2021} informed by existing literature in~\cite{notbohm2016modification,ROSENBLUM2001279,tass1998,Lachaux1999,lowet2016quantifying}. We have tested these measures on the existing LGN network~\cite{mahajan2021}. Subsequently, based on physiological studies~\cite{shermanbook,sherman2001-ciruit}, we have extended this network to include cortical layer 4 (that is known to be the main recipient of the thalamic input), and layer 6 (that is known to be the main source of cortical feedback to the thalamus); the layers 4 and 6 populations project on to one another. The \textit{in silico} model presented here consists of these three interconnected thalamocortical modules. At first, we set the alpha frequency mode in our model, similar to the IAF identification in Notbohm \emph{et al}'s study. The peak alpha frequencies in the responses of the three modules lie between 10 -- 13 Hz. Next, we provide periodic and jittered pulse inputs to the model, superimposed with noise to simulate intrinsic noise introduced during biological retinal processing. We use the objective metrics collated in~\cite{mahajan2021} to measure phase synchrony in the network.

Our results show that for the case of periodic input, the fundamental frequencies in the outputs of all three modules of the thalamocortical network are same as that of the pulse stimulus. Furthermore, time series outputs of all three modules are phase locked when the amplitude of the input is high. However, with jittered input, there is no phase locking even with high input amplitude; also, the frequency domain response does not follow the input frequency. This is in agreement to Notbohm \emph{et al}'s observations in SSVEP. We note that the inter-module connectivities in the model need to be set to relatively low values to simulate the alpha rhythm state. With reduced synaptic connections between the LGN and the cortical layers, as well as between the two cortical layers, the outputs of both cortical layers oscillate in the alpha rhythmic band. However, in this alpha rhythmic state, both cortical layers are unresponsive to periodic stimuli, even with high pulse amplitudes. In contrast, when the inter-module synaptic connectivities are set to relatively higher values, the alpha rhythm disappears, and the cortical layers start responding to periodic stimuli. Thus, our study predicts a role of the thalamocortical circuit connectivities in simulating different brain `states'. In this regard, an experimental study on visual discrimination task by Hanslmayr \emph{et al}~\cite{hanslmayr2005} observed that when human subjects performed poorly in perceptual tasks, their EEG indicated large amplitude alpha oscillations. The authors speculate that such a `state' of the brain occurs when the cortex is largely `deactivated' with fewer synaptic activity. Conversely, a good perceptual task performance correspond to low amplitude alpha oscillations in the EEG. Such a brain state occurs when the cortex is `activated'. Our observations with the \textit{in silico} model are in alignment with these experimental observations: the unresponsive (to periodic flickering stimuli) network state with reduced inter-module connectivity is similar to the state of cortical deactivation, with reduced perception of visual stimuli, and responding with dominant frequency within the alpha rhythm; conversely, with increased inter-module connectivity, the alpha rhythms cease and the model starts responding to visual stimuli, simulating the state of cortical activation.

At this point, we would like to specify a NMM-based \textit{in silico} research where Vindiola \emph{et al}~\cite{vindiola2014} studied simulated EEG data to investigate efficacy of three phase synchronization measures viz.\ phase locking value, imaginary component of the coherence, and debiased weighted phase lag index. Their NMM represented the cortical region consisting of three neuronal populations viz.\ a pyramidal layer, an excitatory spiny stellate layer, and an inhibitory interneuron layer; this was configured to create a dual-kinetic model as described in~\cite{david03}. Twenty-eight experimental conditions were simulated by varying the timeframe (slow or fast oscillations) and frequencies (mono-kinetic with a pure frequency or dual-kinetic with a dominant frequency) of the underlying connectivity pattern of the model. The authors concluded that none of their phase synchronisation measures outperformed another; rather, the efficacy of a certain measure or a combination of the measures is user specific. This is in line with our findings where a combination of PLV and CPI has proved to be more suitable to detect phase synchronisation with changing input conditions. Vindiola \emph{et al} further highlight the promise that neural mass models hold to understand brain activity dynamics; this is also evident through our study conducted here.

In summary, we have studied the population dynamics underpinning phase entrainment of default brain rhythms such as seen during SSVEP using a biologically inspired population \textit{in silico} model. In doing so, we have validated the model with an experimental study. The \textit{in silico} model used in this work is presented in Sections~\ref{sec:211} and~\ref{sec:212}; the simulation design and methods are presented in Sections~\ref{sec:213} --~\ref{sec:215}. The measures of phase synchrony as implemented in this work and the related simulation methods are presented in Section~\ref{sec:22}. Simulation results are presented in Section~\ref{sec:3} and observations are made in context. A detailed discussion of this study is presented in Section~\ref{sec:4}. Conclusion is made in Section~\ref{sec:5}.

\section{Materials and Methods}
\label{sec:2}
\subsection{A Biologically Inspired Thalamocortical Network}
\label{sec:21}
\begin{figure} [b!]
\centering
\includegraphics[scale=0.60]{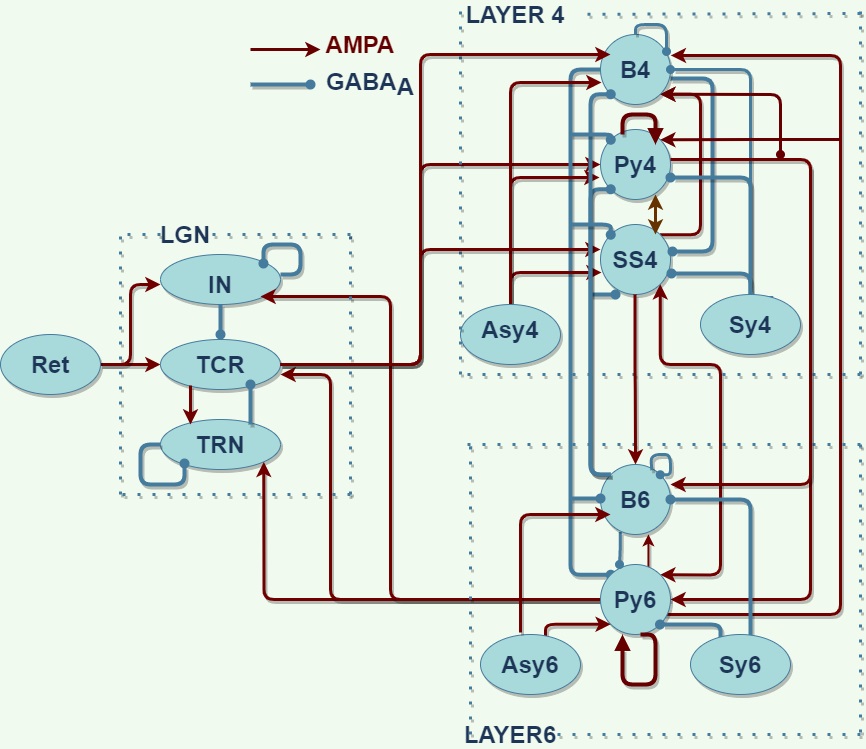}
\caption{The biologically inspired population neural network consists of three modules viz.\ layer 4 (L4), layer 6 (L6) of the visual cortex, and the visual thalamus (Lateral Geniculate Nucleus (LGN)). The LGN neural populations receive inputs from the retina (Ret).The Relay cells (TCR) of LGN project on to the L4 populations. The L4 and L6 cells connect bidirectionally. The L6 populations project back to the LGN, thus forming a closed loop. Both cortical layers are provided with noisy inputs simulating excitatory (Asy4, Asy6) and inhibitory (Sy4, Sy6) projections from other parts of the cortex. The outputs of this \textit{in silico} model are the time series responses of excitatory populations from each of the three modules viz.\ TCR cells of the LGN, Py4 cells of L4, and Py6 cells of L6. The excitatory and inhibitory synaptic connections are shown as AMPA-based and GABA$_A$-based respectively.}
\label{fig:TCT}
\end{figure}
The three module \textit{in silico} thalamocortical network is shown in Fig.~\ref{fig:TCT}. A brief biological background to the model is provided below, followed by the mathematical equations and parameters that define the network.

\subsubsection{Structure and Layout}
\label{sec:211}

The retino-geniculate layout is as in our previous works~\cite{Bhattacharya2011,bhattacharya2016causal}. Retinal spiking neurons (Ret) communicate environmental information to the LGN, which is the thalamic nuclei of the visual pathway. The LGN consists of two cell populations viz.\ the excitatory thalamocortical relay (TCR) cells, and local inhibitory interneurons (IN). A thin sheet of inhibitory tissue `surrounds' the thalamus partially, called the Thalamic Reticular Nucleus (TRN). The main recipients of the retinal inputs are the TCR and IN populations. The TRN population receives `copies' of all feed-forward and feedback communications between the thalamus and the cortex. Because of its integral role in thalamic dynamics, the TRN is considered a part of the thalamus. All structural and connectivity data on LGN are derived from physiological studies~\cite{shermanbook,sherman2001-ciruit}, and are as in our previous works.

The neocortex in mammals is known to have a fairly homogeneous `laminar' structure, consisting of six layers (See~\cite{Bhattacharya2011,sherman2013functional} for detailed references). The main excitatory type in all cortical layers are the Pyramid (Py) cells, and the main inhibitory interneurons are the Basket (B) cells. There are a myriad of other inhibitory interneuron types in all layers of the cortex, and those are often referred collectively as non-Basket cells. In addition, the fourth layer (L4) of the cortex is known to consist of an excitatory interneuron variety called Spiny-Stellate (SS4) cells. Physiological studies show that the topology in the cortex follow a columnar structure, where the six layers (vertical) under a tissue surface area of $\approx1$ mm$^2$ is called a `cortical micro-column'. The whole cortical tissue may be thought of as vertical `stacks' of several microcolumns with intra- and inter-columnar connectivities~\cite{douglas2007mapping}. It is this columnar architecture that makes the cortex popular with \textit{in silico} modellers.

The visual cortex is by far the most commonly studied cortical area, contributing to available physiological data~\cite{sherman2013functional} that provide the substrate for building biologically inspired neural networks. In the visual pathway, the retinal information is further transmitted to the cortex by the TCR cells of the LGN, that project mainly to L4 of the visual cortex. All cells of the LGN receive feedback from the visual cortex, primarily from the Py cells of the L6. For simplicity in our thalamocortical network, we have included only these two layers i.e\ the L4 and L6 modules (see~\cite{shermanbook}). Both modules have an excitatory Py population and an inhibitory B population; in addition, L4 has the SS4 population. The intra- and inter-layer connectivity parameters in L4 and L6 are drawn from physiological study of the cat visual cortex~\cite{binzegger2004quantitative}.

\begin{table} [ht]
\scriptsize
\centering
\caption{Synaptic connectivity parameters for cortical and thalamic populations as derived from \cite{binzegger2004quantitative} and~\cite{bhattacharya2016causal} respectively. All values are normalised as percentage. Parameters marked in `-' are not defined.}
\begin{tabular}{|p{35pt}|p{25pt}|p{30pt}|p{30pt}|p{30pt}|p{30pt}|}
\hline
\multicolumn{6}{|c|}{\textbf{LGN Parameters}} \\
\hline
\textbf{Pre$\rightarrow$ Post$\downarrow$} & \textbf{Ret} & \textbf{TCR} & \textbf{TRN}&\multicolumn{2}{c|}{\textbf{IN}}\\
\hline
TCR& 7.1&-&15.45&\multicolumn{2}{c|}{15.45}\\
\hline
TRN& - &35&20&\multicolumn{2}{c|}{-}\\
\hline
IN& 47.4&-&-&\multicolumn{2}{c|}{23.6}\\
\hline
\multicolumn{6}{|c|}{\textbf{L6 Parameters}} \\
\hline
\textbf{Pre$\rightarrow$ Post$\downarrow$} & \textbf{Py6} & \textbf{B6} & \textbf{Asy6}&\multicolumn{2}{c|}{\textbf{Sy6}}\\
\hline
Py6&4.1&9.6&26.4&\multicolumn{2}{c|}{5.72}\\
\hline
B6&4&11.4&25.9&\multicolumn{2}{c|}{5.61}\\
\hline
\multicolumn{6}{|c|}{\textbf{L4 Parameters}} \\
\hline
\textbf{Pre$\rightarrow$ Post$\downarrow$} & \textbf{Py4} & \textbf{B4} & \textbf{SS4}& \textbf{Asy4} &\textbf{Sy4} \\
\hline
Py4&4.1&9.6&15.45&26.4&5.72 \\
\hline
B4 &4&11.4&15.2&25.9&5.61 \\
\hline
SS4&4.1&9.4&15.45&26.5&5.74 \\
\hline
\end{tabular}
\label{tab:model_params}
\end{table}
The nomenclature for the cortical layer populations adopted in this work are as follows: L4 populations are Py4, SS4, B4; L6 populations are Py6, B6; sources of intrinsic noisy input from other cortical and subcortical areas to L4 (L6) are Asy4 (Asy6) (asymmetric: excitatory) and Sy4 (Sy6) (symmetric: inhibitory). The network outputs are the time series responses of the TCR, Py4 and Py6 populations, all of which are known to communicate neuronal information over long distances due to their physical attributes, as opposed to the interneurons that are known to communicate locally. Intra-module synaptic connectivity parameters are mentioned in Table~\ref{tab:model_params}. Inter-module connectivity parameters are a part of our simulation methods described in Sections~\ref{sec:214}~--~\ref{sec:215}. The mathematical framework underlying the \textit{in silico} model is described in the following Section~\ref{sec:212}.
\subsubsection{Model equations}
\label{sec:212}
Each synaptic activity in the population neural network is governed by~(\ref{T_eq})~--~(\ref{V_eq})~\cite{bhattacharya2016causal}:
\begin{itemize}
    \item The concentration of the neurotransmitter ($[T(V_{pre})]$) released at the synapse is a function of the voltage of the presynaptic neuron ($V_{pre}(t)$) and is defined by a sigmoid function:
    \begin{equation} \scriptsize \label{T_eq} 
        T(V_{pre}(t)) = \frac{T_{max}}{1 + e^{-\frac{(V_{pre}(t) - V_{thr})}{\sigma}}}
    \end{equation}
where $T_{max}$ is the maximum neurotransmitter concentration approximated by 1 milliMole; the parameter $V_{thr}$ represent the threshold at which $T=0.5T_{max}$; the $\sigma$ affects the steepness of the sigmoid.

    \item The proportion of the open ion channels ($r$) on the postsynaptic ensemble membrane due to the binding of neurotransmitters with the neurotransmitter-receptors is defined as: 
    \begin{equation} \scriptsize \label{r_eq}
        \frac{dr(t)}{dt} = \alpha\;T\;(V_{pre}(t))\;(1 - r(t)) - \beta\; r(t)
    \end{equation}
    where $\alpha$ and $\beta$ are forward and reverse rates of chemical reaction respectively.
    \item The current in the postsynaptic population ($I_{psc}(t)$) due to the opening of ion channels is defined as:
    \begin{equation}  \scriptsize \label{I_eq}
        I_{psc}(t) = C_{uv}\;g_{max}^{syn}\;r(t)\;(V_{psp}(t) - E_{rev}^{syn})
    \end{equation}
where $C_{uv}$ refers to the connectivity parameter of the synapse from presynaptic population $u$ to postsynaptic population $v$ (see the pre and post populations in Table~\ref{tab:model_params}); Each connectivity parameter is either excitatory (AMPA) or inhibitory (GABA$_A$) synapse (See Fig.~\ref{fig:TCT} for these details); $V_{psp}(t)$ is the postsynaptic voltage, $g_{max}^{syn}$ is the maximum conductance of the postsynaptic ensemble membrane potential due to opening of ion channels, where $syn \in \{AMPA \quad GABA_A\}$, $E_{rev}^{syn}$ is the reverse potential of $syn$. 
   \item The ensemble postsynaptic membrane potential is the sum of the currents in the postsynaptic population due to converging synapses and is defined as:
    \begin{equation} \scriptsize \label{leak_eq}
        \kappa \frac{dV_{psp}(t)}{dt} = -\sum_{uv} (I_{psc}(t)) + I^{leak}(t)
    \end{equation}

    where $\kappa$ is the membrane Capacitance (pF) and the parameter $I^{leak}$ is the ensemble leak current of the postsynaptic population membrane and given as:
    
    \begin{equation} \scriptsize \label{V_eq}
        I^{leak}(t) = g^{leak}(V_{psp}(t) - E^{leak})
    \end{equation}
    
    Here $g^{leak}$ and $E^{leak}$ are the maximum leak conductance and leak reversal potential respectively of the postsynaptic cell population.
\end{itemize}

\begin{table}
\scriptsize
\centering
\caption{Synaptic connectivity parameters for each neuron population in the thalamocortical neural network modelled with ~(\ref{T_eq})~--~(\ref{V_eq}). }
\begin{tabular}{|p{100pt}|p{30pt}|p{30pt}|p{25pt}|p{30pt}|p{25pt}|p{25pt}|p{25pt}|p{25pt}|p{25pt}|}
\hline
\multicolumn{10}{|c|}{\textbf{(A) Neurotransmission   Parameters}} \\
\hline
\textbf{Parameters} &\textbf{Value} & \multicolumn{8}{c|}{\textbf{Synaptic Pathway}}
 \\
\hline
{$\alpha((mM)^{-1}.(s)^{-1})$} &100 & 
\multicolumn{8}{c|}{AMPA,GABA$_{A}$}  \\
\hline
{$\beta(s^{-1})$}&50&\multicolumn{8}{c|}{AMPA}\\
   \cline{2-10}&50&\multicolumn{8}{c|}{AMPA} \\
   \cline{2-10}&40&\multicolumn{8}{c|}{GABA$_{A}$} \\

\hline
{$g^{\bar{\eta}} (\mu S/cm^{2})$} & 1000 & \multicolumn{8}{c|}{GABA$_{A}$}  \\
   \cline{2-10}&300&\multicolumn{8}{c|}{AMPA(Ret to TCR)} \\
   \cline{2-10}&100&\multicolumn{8}{c|}{AMPA(Ret to IN) (TCR to TRN)} \\
\hline
{$E^{\bar{\eta}}   (mV)$} & 0 & \multicolumn{8}{c|}{AMPA}  \\
   \cline{2-10}&-85&\multicolumn{8}{c|}{GABA$_{A}$(inter-population)} \\
   \cline{2-10}&-75&\multicolumn{8}{c|}{GABA$_{A}$(recurrent)} \\
\hline
\end{tabular}\\

\begin{tabular}{|p{70pt}|p{16pt}|p{24pt}|p{25pt}|p{25pt}|p{24pt}|p{25pt}|p{21pt}|p{20pt}|p{20pt}|}
\hline
    \multicolumn{10}{|c|}{\textbf{(B) Cell Membrane Parameters}} \\
\hline
\textbf{Parameters} & \textbf{Ret} & \textbf{TCR} &
\textbf{IN} &
\textbf{TRN} &
\textbf{Py4
Py6} &
\textbf{B4
B6} &
\textbf{SS4}&
\textbf{Asy} &
\textbf{Sy} 
 \\
\hline
{$g^{leak} (\mu S/cm^{2})$} & X   & 10  & 10    & 10    & 10  & 10    & 10  & X   & X   \\
\hline
{$E^{leak}   (mV)$} & X   & -55 & -72.5 & -72.5 & -55 & -72.5 & -55 & X   & X   \\ \hline
$V_{rest}   (mV)$     & -65 & -65 & -75   & -85   & -65 & -85   & -65 & -65 & -75 \\ \hline

\end{tabular}

\label{tab:syn_params}
\end{table}
Assuming that AMPA- and GABA$_A$-based neurotransmission have similar dynamics in the thalamus and the cortex, as well as due to unavailability of exact physiological data, the neurotransmission parameters for the cortical excitatory populations viz. Py4, Py6, SS4 are set identical to those of the excitatory population of the LGN viz.\ TCR. Similarly, neurotransmission parameters for the cortical inhibitory cells B4 and B6, are set according to the inhibitory LGN cells. All parameters used in the model equations are mentioned in Table~\ref{tab:syn_params} and are as in~\cite{bhattacharya2016causal}. 

\subsubsection{Simulation methods}
\label{sec:213}
Consistent with the identification of IAF in~\cite{notbohm2016modification} prior to studying the SSVEP, we start by simulating in our \textit{in silico} model the condition of alpha rhythm corresponding to awake resting state with eyes closed. Towards this, the retinal input to the LGN is a random noise, which is consistent with all previous alpha rhythm models. Next, to simulate retinal response to flickering visual stimuli, three different types of inputs are provided to the model, which is consistent with the experimental set-up of Notbohm \emph{et al}~\cite{notbohm2016modification}. The details are specified below:
\begin{itemize}
\item Noise: The excitatory noise inputs to the network populations viz.\ Ret to LGN populations TCR and IN, Asy4 to L4 populations, and Asy6 to L6 populations, are simulated with random noise in MATLAB, normalised to a mean -65 mV and standard deviation 2 mV. The inhibitory sources of noise to both cortical layers viz.\ Sy4 and Sy6 are random noise with mean -75 mV and standard deviation 2 mV. 
\item Periodic flicker: This is a pulse train with constant inter-pulse interval $T$, and therefore constant frequency $f=1/T$. Frequency is varied between 1 -- 30 Hz, and pulse amplitudes are varied from 1 -- 10 mV. The pulse train `on-time' $p_{on}$ in this work is maintained at 1 ms, which is one simulation time-step. Note that the input to the network is assumed to be from the retina, i.e.\ processed by the retina, which are trains of spikes from retinal spiking neurons. Thus, we keep the $p_{on} \rightarrow 0$ (time units) to simulate ensemble membrane potential of the retinal spiking neurons corresponding to periodic pulse stimuli. To simulate intrinsic brain noise that is added during retinal processing, the pulse train is superimposed with the aforementioned random noise.
\item Jittered flicker: To generate the jittered signals similar to that described in~\cite{notbohm2016modification}, the inter-pulse intervals are drawn from a uniform distribution with  maximum bounds of $\pm 60\% \; of \; (\frac{1}{f_1})$, where $f_1$ is our desired jittered pulse train frequency. For example to generate a jittered pulse train with $f_1=10$ Hz, we do the following: (a) calculate $\tau = \frac{1000  ms}{f_1} = 100$ ms; (b) generate a random vector $S \in {\pm60\% \; of \;  (\frac{1}{f_1}) } \Rightarrow 40$ ms $\leqslant S \leqslant 160$ ms. The inter-pulse-interval in the jittered signal with 10 pulses will now be drawn randomly from $S$. Finally, for reasons aforementioned in the case of periodic pulse train input, this jittered pulse train is superimposed with random noise. All other pulse and noise attributes are similar to that for periodic pulse input.
    \item Mixed signal input: In Notbhom \emph{et al}'s experiment, $5 \times 7$ intensity-frequency combinations were used for providing an input stimulus that had a random mix of periodic, jitter and noise signals; `intensity' refers to the amplitude of the driving force. For simplicity in this work, we use six non-repetitive frequencies drawn randomly from a sample of 10~--~15 Hz (increments of 1 Hz) while keeping the amplitude constant at 10 mv. Thus, we have a smaller subset of $1 \times 6$ amplitude-frequency combinations to work with. Each condition (amplitude-frequency combination) is simulated for 30 s with jittered signal, followed by 30 s with periodic signal, and finally for 30 s with noise input. The resulting continuous signal of 540 s $ ((30 \times 3) \times (1 \times 6))$ is fed to the network, simulating continuous visual stimuli with varying input conditions.
\end{itemize}
All simulations are done on MATLAB (version 2020a). Simulation time of the network (except for the mixed signal input) is $T=120$ s at a resolution of $\Delta t=1 $ ms. The final time series (membrane potential) for each output population viz.\ TCR, Py4, Py6 are averaged over 20 trials, where each trial is simulated with separate noise (or additive noise in the case of pulse stimuli) inputs. Frequency response of the average membrane potential is computed as power spectral density of the response time series using Welch periodogram, at sampling frequency of $\frac{1}{\Delta t}=1000$ Hz and 50\% overlap. To understand time-frequency response in nonstationary responses with respect to jittered flicker and mixed signal input, we computed the short-time-Fourier-Transform (STFT) with time window length of 1000 ms and 50\% overlap of the consecutive windows. Simulation methods for quantitative measures of phase synchrony are discussed in Section~\ref{sec:22}. In the following sections, we explain the methods for parameterising the model to simulate resting state alpha rhythm and SSVEP-like responses in the model.
\subsubsection{Parameterising for Alpha Rhythm}
\label{sec:214}
As mentioned in Section~\ref{sec:211}, the inter-module connectivity parameters are a part of our simulation methods as we did not find any physiological reference for these in existing literature. We set the inter-module parameters heuristically such that the response of each output population viz.\ TCR, Py4 and Py6 have peak power within the alpha frequency band. In addition, the afferent connection weights from Asy4 and Asy6 are reduced from their base values given in Table~\ref{tab:model_params}. The parameters specific to this alpha rhythmic state of the model are mentioned in Table~\ref{tab:intermod_conn_params_noise}. This exercise of finding the peak alpha frequency for each of the three model responses may be thought to be equivalent to finding the IAF in~\cite{notbohm2016modification} prior to studying phase synchrony in SSVEP.

Readers may note that in addition to the inter-module connectivity, the inhibitory projections to the TCR in the LGN module are also altered compared to our base values in Table 1, and is based on our previous observations in an \textit{in silico} model of the LGN~\cite{bhattacharya2016causal}. The total inhibitory projections from both IN and TRN is reported as 30.9\% of the total afferent synapses to the TCR in physiological studies (see~\cite{Bhattacharya2011} for a discussion). However, the exact proportion of the synapses from IN and TRN are not known, and are treated as model hyper-parameters. In our previous works, we observed that increasing the inhibitory connection weight of the TRN to TCR pathway, compared to that of the IN to TCR pathway, was conducive to generating alpha rhythms in the model. Conversely, an increased efficacy in retinal transmission was observed when the connection weight of the IN to TCR pathway was greater than that of the TRN to TCR pathway. We adjust the proportions of inhibitory synapses from the IN and TRN on to the TCR populations accordingly (compare Tables~\ref{tab:intermod_conn_params_noise} and~\ref{tab:intermod_conn_params_ssvep}). In Section~\ref{sec:31}, we demonstrate the alpha rhythmic response of the network in the `deactivated' cortical state.
\begin{table}
\scriptsize
\centering
\setlength{\tabcolsep}{3pt}
\caption{Inter-module connectivity parameters to simulate `deactivated' cortical state where the dominant frequency of oscillation for all network output populations are within the alpha band (8~--~13 Hz) corresponding to noise input. Parameters set to: `*' are same as in Table~\ref{tab:model_params};`-' are not defined. }
\begin{tabular}{|p{30pt}|p{30pt}|p{30pt}|p{25pt}|p{25pt}|p{25pt}|p{25pt}|p{25pt}|p{25pt}|p{20pt}|p{20pt}|}
\hline
\textbf{Pre$\rightarrow$ Post$\downarrow$}&\textbf{TCR}&\textbf{TRN}&\textbf{IN}& \textbf{Py4}&\textbf{SS4}&\textbf{B4}&\textbf{B6}&\textbf{Py6}&\textbf{Ay}&\textbf{Sy}\\
\hline
TCR&-&23.17&7.73&-&-&-&-&1&-&- \\
\hline
TRN&*&*&-&-&-&-&-&1&-&- \\
\hline
IN&-&-&*&-&-&-&-&1&-&- \\
\hline
Py4&10&-&-&*&*&*&9.6&11.4&5&* \\
\hline
SS4&10&-&-&*&*&*&9.4&11.4&5&* \\
\hline
B4&11.4&-&-&*&*&*&11.4&11.4&5&* \\
\hline
B6&-&-&-&12&12&10&*&*&5&* \\
\hline
Py6& -&-&-&10&10&8.5&*&*&5&* \\
\hline
\end{tabular}

\label{tab:intermod_conn_params_noise}
\end{table}
\subsubsection{Parameterising for flickering visual stimuli}
\label{sec:215}
To simulate SSVEP-like signals in the network, we provide periodic pulse stimuli as retinal input to TCR. At first, the model is maintained at its alpha rhythmic state as defined in Table~\ref{tab:intermod_conn_params_noise}. Both Py4 and Py6 show a strong response around their respective peak alpha rhythmic frequency, but fail to respond to frequencies $\lessapprox \pm3$ Hz on either side of the peak. Unlike the cortical populations, the TCR population, being a direct recipient of the retinal input, respond with a periodic output at the fundamental frequency of the retinal input, and across all input frequencies. We note that the highest power within the fundamental frequency of the TCR is when the input frequency matches the TCR's alpha rhythmic peak frequency. We speculate that this lack of response in the cortical populations (i.e.\ outside of respective alpha peak frequencies) is due to the reduced network connectivity in the deactivated state of the network. To simulate an `activated' network state, we increase the connection strengths to and from the cortical layers. Thus, connection strengths in the closed loop pathway formed by $TCR\rightarrow Py4 \rightarrow Py6 \rightarrow TCR$ are set to higher values than those mentioned in Table~\ref{tab:intermod_conn_params_noise}. The afferent connections from other parts of the cortex (viz.\ Asy4 and Asy6) are set back to their base values indicated in Table~\ref{tab:model_params}. In addition, the proportion of inhibitory projections to the TCR in the LGN module are adjusted informed by our previous works~\cite{bhattacharya2016causal} as elucidated above in Section~\ref{sec:214}. Table~\ref{tab:intermod_conn_params_ssvep} shows all the parameter values corresponding to the `activated' cortical state of the network. The cortical layers now respond to periodic pulse stimuli. Moreover, corresponding to noise input from Ret to the LGN such as provided during alpha rhythm simulation, the frequency response of the cortical outputs shows a $\frac{1}{f}$ characteristic and no alpha rhythmic peak. This is similar to the frequency response of EEG recorded from an active brain state that is known to show a $\frac{1}{f}$ characteristic~\cite{pritchard92,niedermeyer97,buzsaki2005}. We discuss model response to flickering visual stimuli in Section~\ref{sec:32}.
\begin{table}
\scriptsize
\centering
\setlength{\tabcolsep}{3pt}
\caption{Inter-module connectivity parameters to simulate `activated' cortical state to process flickering visual stimuli. Parameters set to: `*' are same as in ~\ref{tab:model_params};`-' are not defined;`bold font' are the altered values with respect to alpha rhythmic model given in Table~\ref{tab:intermod_conn_params_noise}. }
\begin{tabular}{|p{30pt}|p{30pt}|p{30pt}|p{28pt}|p{25pt}|p{25pt}|p{25pt}|p{25pt}|p{25pt}|p{20pt}|p{20pt}|}
\hline
\textbf{Pre$\rightarrow$ Post$\downarrow$}&\textbf{TCR}&\textbf{TRN}&\textbf{IN}& \textbf{Py4}&\textbf{SS4}&\textbf{B4}&\textbf{B6}&\textbf{Py6}&\textbf{Ay}&\textbf{Sy}\\
\hline
TCR&-&\textbf{7.73}&\textbf{23.17}&-&-&-&-&\textbf{60}&-&- \\
\hline
TRN&*&*&-&-&-&-&-&\textbf{10}&-&- \\
\hline
IN&-&-&*&-&-&-&-&\textbf{30}&-&- \\
\hline
Py4&\textbf{60}&-&-&*&*&*&*&11.4&*&* \\
\hline
SS4&10&-&-&*&*&*&*&11.4&*&* \\
\hline
B4&11.4&-&-&*&*&*&11.4&11.4&*&* \\
\hline
B6&-&-&-&12&12&10&*&*&*&* \\
\hline
Py6&-&-&-&\textbf{60}&10&8.5&*&*&*&* \\
\hline
\end{tabular}

\label{tab:intermod_conn_params_ssvep}
\end{table}
\subsection{Quantifying phase synchronisation}
\label{sec:22}
To quantify synchronisation among neuronal populations in our thalamocortical network, we have taken inspiration from several researches that apply phase locking measures to understand synchrony in brain signals~\cite{ROSENBLUM2001279,tass1998,Lachaux1999,lowet2016quantifying}. Recently, we have collated these measures as a set of objective metrics to measure phase synchronisation in \textit{in silico} models~\cite{mahajan2021}. In this section, we present a brief overview of three objective measures that we have implemented in this work to measure phase synchrony, along with our simulation methods.
\subsubsection{Phase Locking Value (PLV)}
 In~\cite{tass1998}, the authors define synchronisation between two oscillators $x$ and $y$ in terms of their phase locking, where the locking condition is defined as follows:
\begin{equation}
\scriptsize
\mathopen|n \theta_x (t) - m \theta_y (t)\mathclose| < \epsilon \; \forall t  \label{eq:plv1}
\end{equation}
In~(\ref{eq:plv1}), $\theta_{x,y} (t)$ represent the instantaneous phase at time $t$ for the two oscillators, $\epsilon \in \mathbb{R}$ is an arbitrary constant, $n,m \in \mathbb{N}$ represent the ratio of the frequencies of the respective oscillators.

The signal amplitudes are trivial here as they do not affect the phases. Thus, the phase relation can be mapped on the unit circle in the complex plane, see~(\ref{eq:plv2}). The Phase Locking Value (PLV) is defined as the mean of the instantaneous phase differences~\cite{Lachaux1999,lowet2016quantifying}:
\begin{equation}
\scriptsize
PLV = |\frac{1}{T}\sum e^{i(n \theta_x (t) - m \theta_y (t)}|,  \label{eq:plv2}
\end{equation}
where $T$ is the total simulation time, and all other variables are the same as in~(\ref{eq:plv1}). Readers may refer to~\cite{lowet2016quantifying} for a demonstration of the basic concepts underlying phase locking using simple examples.

\textit{Simulation Methods:} All simulations are performed using MATLAB (version 2020a). The phase at each time instant of the noisy input and output signals are extracted using the Hilbert Transform (HT) function~\cite{tass1998,ROSENBLUM2001279}. In this work, we consider only $1:1$ phase relationships in the network i.e.\ $m=n=1$. This phase information is used for all visualisations of phase locking described below:

In MATLAB, the output of the HT function is in rectangular form. This is converted to polar form to plot the \textit{polar histograms} with equally spaced bins. These plots indicate the spread of the phases on the quadrants~---~a peaky cluster within one quadrant implies higher phase locking; conversely, a uniform spread over all quadrants imply a noisy phase relation and absence of phase locking.

The phases for each signal extracted using the HT are used to calculate the phase difference between any two signals. Using the \textit{unwrap} function in MATLAB gives a cumulative phase vector that is plotted against time. This phase vs time plot shows the overall phase locking behaviour between the two signals~\cite{tass1998,notbohm2016modification}, and are referred to as the \textit{phase slip} plots. A high phase locking will have a phase difference that is constant over time indicated by flatter plot. Noisy signals will have a steeper gradient as the cumulative phase difference grows with time.

\textit{Arnold tongues} are often used to understand the phase-locked zones of an oscillator as the frequency and amplitude of its periodic input vary~\cite{glass2001synchronization}. The phase-locked regions grow wider towards the top of the Y-axis (increased input signal amplitude) thus forming a `tongue' shaped (inverted triangle) region; hence the nomenclature. As in~\cite{notbohm2016modification}, we have used the PLVs calculated with varying frequency and amplitude for both periodic and jittered flicker inputs to plot the Arnold tongues.
\subsubsection{Normalised Shannon Entropy (NSE)}
Another measure of phase synchronisation used in~\cite{tass1998,ROSENBLUM2001279,notbohm2016modification} is the entropy in the phase difference between two signals. The concept of entropy was proposed (in 1948) by Claude Shannon in Information Theory , and is often referred to as the Shannon entropy. If a random variable $X_n$ follows a uniform distribution where each event $x_k \in X_n$ occurs with a probability $p_k$, then `information' is defined as $I(x_k) = log(\frac{1}{p_k})$, and Shannon entropy H is defined as the Expected information in the distribution $X_n$ ($E[I(X_n)]$): 
\begin{equation}
\scriptsize
    H(X_n) = \sum_{k=1}^n p_k log(\frac{1}{p_k})
    \label{eq:nse1}
\end{equation}
The maximum entropy in the distribution is $H_{max}(X_n)=log(\frac{1}{N})$ i.e. when every event is equiprobable and N is the length of $X_n$. The normalised Shannon entropy (NSE) is calculated as:
\begin{equation}
\scriptsize
   \rho_{norm}  = \frac{H_{max}-H(X_n)}{H_{max}}
   \label{eq:nse2}
\end{equation}
where $\rho_{norm} \in [0, 1]$ with $0$ representing maximum entropy and a uniform distribution. This normalised measure $\rho_{norm}$ is used as a phase synchronisation index in~\cite{tass1998,ROSENBLUM2001279,notbohm2016modification}.

\textit{Simulation Methods}: In the context of our work, the phase difference distributions (discussed in the previous section) are plotted as histogram with $N=80$ bins~\cite{notbohm2016modification} and using the `probability' option in MATLAB, where the height of each bin represents the probability of all phase difference values that fall within that bin. The entropy of this distribution is calculated using~(\ref{eq:nse1}) and normalised using~(\ref{eq:nse2}) to obtain the NSE synchronisation index $\rho_{norm}$, where $\rho_{norm} \rightarrow 1$ corresponds to a peaky distribution of the phase differences implying phase synchronisation. 
\subsubsection{Conditional Probability Index (CPI)}
A third synchronisation index implemented in~\cite{tass1998,ROSENBLUM2001279} based on conditional probability is reported to be a good measure to detect weak phase interactions between two signals. Let $\theta_x(t_j)$ and $\theta_y(t_j)$ be the instantaneous phases of two signals $x(t)$ and $y(t)$ at any time instant $t_j \in t$. Let the phases of each signal be distributed into N time bins. Then, the conditional probability of $\theta_x(t_j)$ being in a certain bin $l \in[1, N]$ when $\theta_y(t_j) \in l$ is defined in~(\ref{eq:cpi1}):
\begin{equation}
\scriptsize
    r_l(t_j)=\frac{1}{M_l}\sum e^{i\theta_x(t_j)} \, \forall j
    \label{eq:cpi1}
\end{equation}
where $M_l$ is the total number of points in the bin $l$~\cite{tass1998}, $r_l(t_j) \in [0, 1]$ with $0$ indicating no dependencies between the two phases. The conditional probability index (CPI) is computed as the average of the conditional probabilities over all the $N$ bins and is defined in~(\ref{eq:cpi2})~\cite{tass1998}. A high CPI implies strong phase synchronisation.
\begin{equation}
\scriptsize
    \lambda_{x,y} = \frac{1}{N}\sum_{l=1}^N r_l(t_j)
    \label{eq:cpi2}
\end{equation}

\textit{Simulation Methods:} We use this third measure in our work for two reasons: First, as a way of confirming the metrics indicated by PLV and NSE. Second, all input signals to our network have additive noise, and we would like our objective metric to detect phase information that may be weakened by the presence of noise. The phases of both signals extracted using the HT (when calculating PLV) are now spread into a histogram with $N = e^{0.626 + 0.4log(M - 1)}$ bins, which is said to be a good measure in~\cite{ROSENBLUM2001279}, $M = \frac{T}{\Delta t}$ is the total number of data points of simulation over a period $T=120$ s at resolution $\Delta t=1$ ms. Note that this synchronisation index was not implemented in~\cite{notbohm2016modification}. As discussed above, we consider only $1:1$ phase synchronisation, and have ignored $m$ and $n$ in~(\ref{eq:cpi1})~--~(\ref{eq:cpi2}).

Overall, the phase synchronisation measures, combined with time series and frequency domain analysis is used to understand interplay of population dynamics in the network corresponding to combinations of simulated input stimuli as well as parameter settings. The results are discussed below.

\section{Results}
\label{sec:3}
\subsection{Simulating Alpha Rhythm}
\label{sec:31}

\begin{figure} [t!]
\begin{center}
\includegraphics[scale=0.78]{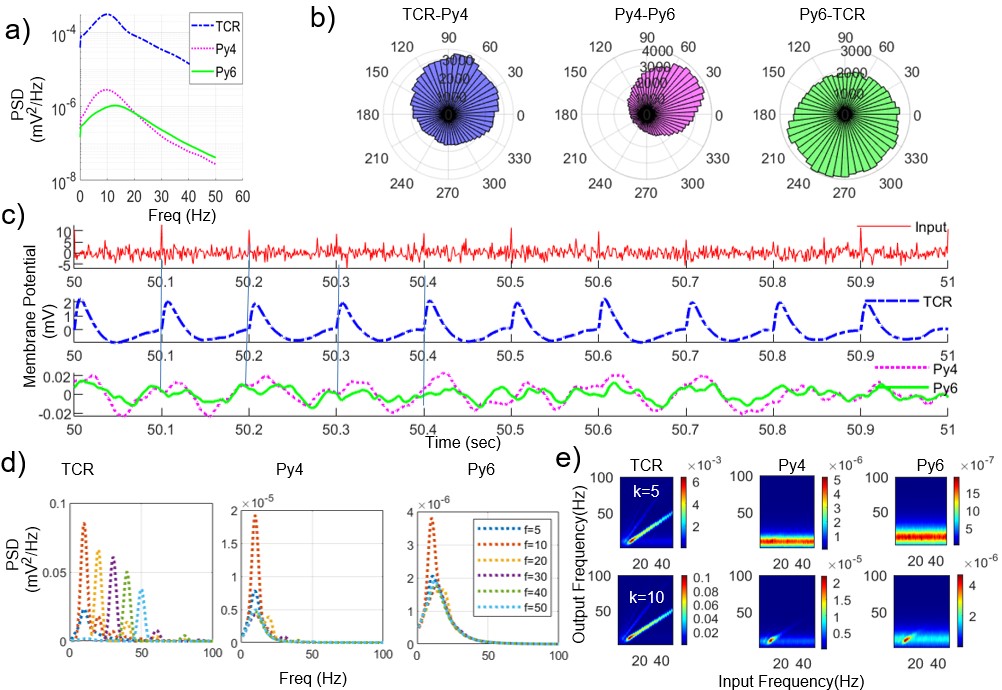}
\caption{(a) The peak within the alpha frequency band ($\alpha_{peak}$) for the three excitatory populations in the thalamocortical network viz.\ the TCR population of the LGN, the Py4 and Py6 populations of the cortical layers 4 and 6 respectively. (b) Polar histograms show a uniform spread of the phase relations between the time series of the three populations corresponding to noise input, and when all three populations oscillate at their respective $\alpha_{peak}$. (c) (top) Time series of the input pulse train overlaid with noise provided as Retinal input; (middle) the TCR response is aligned in phase and frequency with the input stimulus; however (bottom) the Py4 and Py6 populations fail to respond to the periodic visual input. Here, the mean membrane potential of the populations is baseline corrected for comparative scales. The (d) line plots of the power spectral density of Py4 and Py6 show the lack of response to all frequencies except at and around their respective $\alpha_{peak}$. However, the TCR responds in alignment with the input, being a direct recipient of the Ret population. (e) Herrmann-like plots further agree with our observations with the line plots over a wider frequency range, and with increasing pulse amplitude. } 
\label{fig:2}
\end{center}
\end{figure}
Fig.~\ref{fig:2}a shows the power spectral density (PSD) of the three output populations. The peak alpha frequency for TCR and Py4 is between $10~-~11$ Hz; for Py6, it is $\approx 13$ Hz. The model input is noise, and the inter-module connectivity parameters are set to relatively lower values (Table~\ref{tab:intermod_conn_params_noise}) to simulate the state of cortical deactivation. The polar histogram plots in Fig.~\ref{fig:2}b indicate the noisy phase distribution in the three populations. Periodic pulse input with frequency varying between 1 and 50 Hz is provided to the model in this alpha rhythmic state. A sample time series response for pulse input of 10 Hz and amplitude 10 mV, superimposed with noise is shown in Fig.~\ref{fig:2}c. A qualitative observation indicates that the phases of the pulse responses in the output time series of the TCR (being a direct recipient from the Ret population) are aligned with the phases of the corresponding input pulses unlike the Py4 and Py6. Readers may note, the membrane potentials of the populations is baseline corrected by subtracting their mean. The actual membrane potential values are TCR = -70.3, Py4 = -56.6 and Py6 = -59.2 mV. The corresponding line plots of the PSD in Fig.~\ref{fig:2}d shows the TCR output is following the input frequencies. However, the only response in the Py4 and Py6 population outputs are at their respective peak frequencies within the alpha band ($\alpha_{peak}$) that was observed with noise input, regardless of the input frequency. Furthermore, the power amplitude for frequencies $\approx \pm \alpha_{peak}$ is about twice of that for other frequencies.

In our previous work~\cite{Labecki2016}, we have used a visualisation that allow the observation of model outputs over a wider range of frequencies and at a finer resolution. This type of visualisation was originally used by Herrmann \emph{et al} in their study of human EEG~\cite{Herrmann2001}. Other \textit{in silico} studies have also used this visualisation~\cite{robinson2015}. To make a detailed study of the  `deactivated state' of our model, we vary the input frequencies between 1 to 50 Hz at a resolution of 1 Hz. Further, we run several epochs of frequency variation as we increase the pulse amplitude $k$ from 5 to 15 mV at a resolution of 1 mV. The `Herrmann-like' plots are shown in Fig.~\ref{fig:2}e for $k=5, 10$. Similar to what we observed in the line plots, the Py4 and Py6 respond only at their respective $\alpha_{peak}$ regardless of the input frequencies. For $k=10$, there is some response for other input frequencies but only in the vicinity of their respective $\alpha_{peak}$.
\subsection{Simulating Entrainment with Periodic Input}
\label{sec:32}

\begin{figure} [!b]
\begin{center}
\includegraphics[scale=0.67]{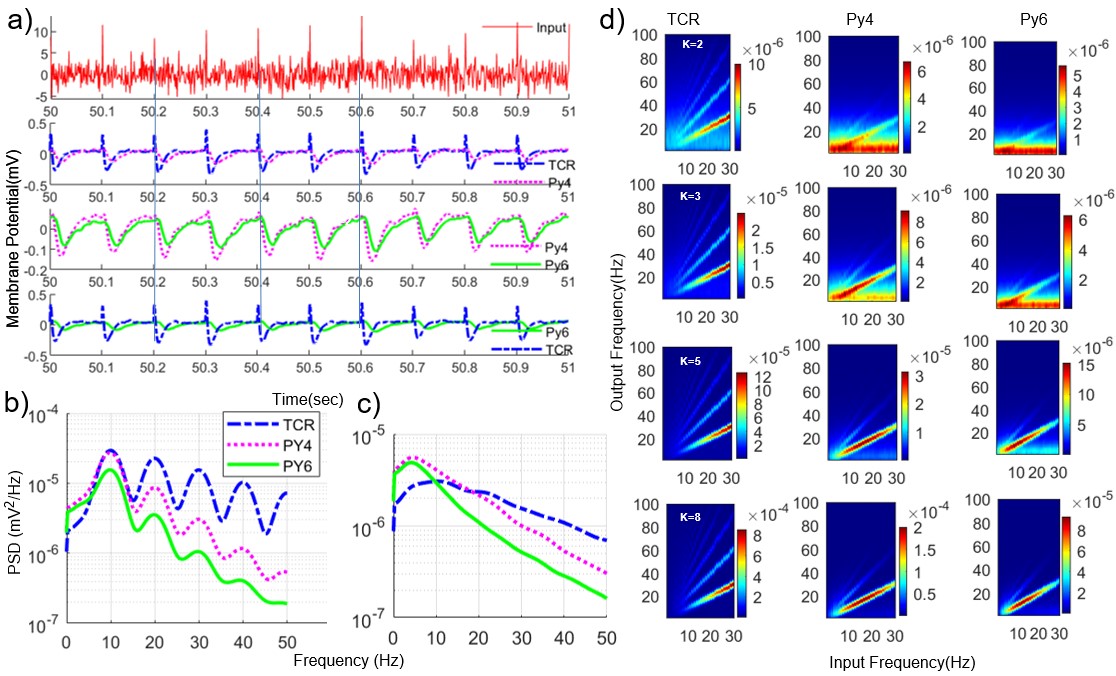}
\caption{(a) Time series  and (b) frequency domain responses of the network when parameterised to simulate an `activated' cortical state, and provided with periodic flicker input of 10 Hz in (a) and (b), while uniform noise input in (c). A pairwise comparison between the output population time series in (a) demonstrate qualitatively the phase locking between the populations, as well as with the periodic pulse input. Here, the mean membrane potential of the populations is baseline corrected for comparative scales. The $1/f$ characteristic in (c) indicating the underlying low power noisy output. (d) Herrmann-like plots illustrate the observations with the line plots over a wider frequency range, and with increasing pulse amplitude. Here, we see that both cortical modules now respond to periodic stimuli.}
\label{fig:3}
\end{center}
\end{figure}

\begin{figure} [hb!]
\centering
\includegraphics[scale=0.90]{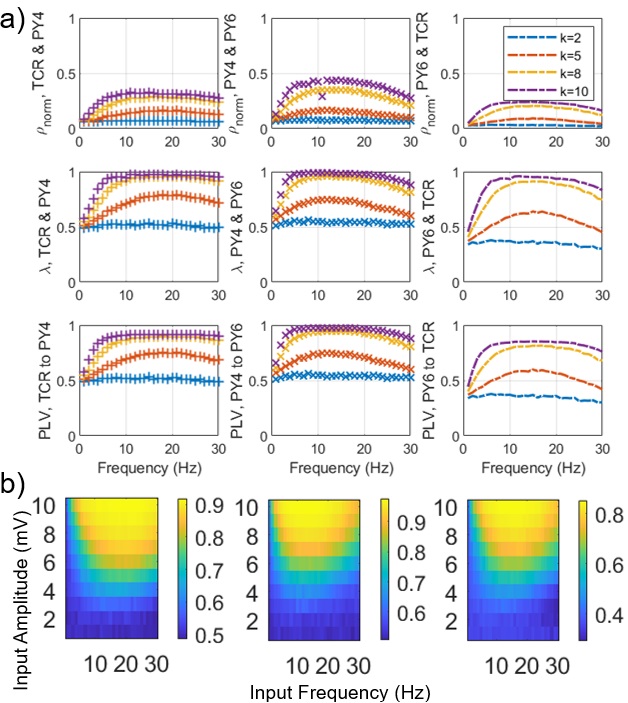}
\caption{(a) Normalised Shannon entropy (NSE: $\rho_{norm}$), Conditional Probability Index (CPI: $\lambda$) and Phase Locking Value (PLV) show the strength of phase synchrony between the output populations viz.\ TCR, Py4 and Py6. All measures are for pulse input frequency $f=10$ Hz, and amplitude $k=2, 5, 8, 10$ mV. The synchronisation increases with increasing $k$. (b) Arnold Tongue formation indicate greater phase synchronisation with higher $k$ (Y-axix).}
\label{fig:4}
\end{figure}

\begin{figure} [hb!]
\centering
\includegraphics[scale=0.90]{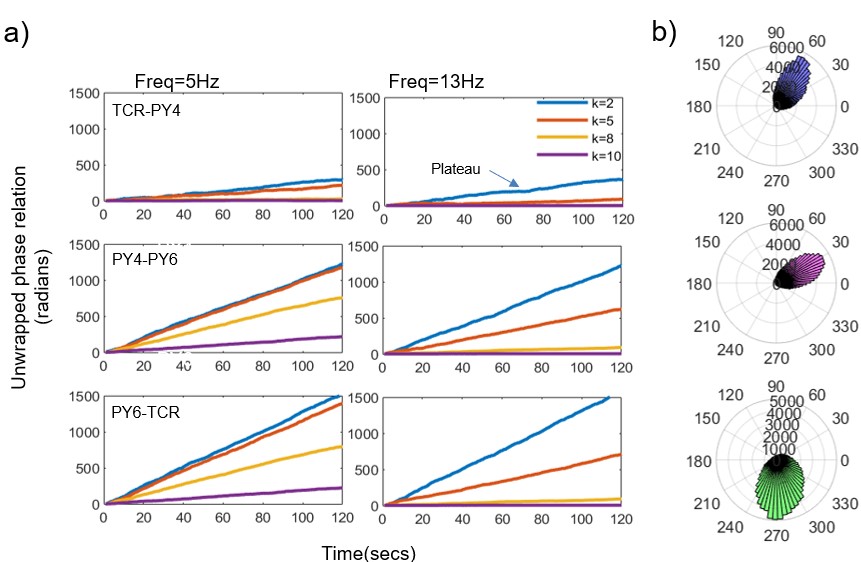}
\caption{(a) Phase slip plots demonstrating the strength of phase synchrony between the output populations viz.\ TCR, Py4 and Py6. The phase synchrony increases with increasing pulse amplitudes $k=2, 5, 8, 10$ mV as well as with increasing frequency $f=5, 10$ Hz, when the plots get flatter. (b) polar histograms for $k=5$ mV and $f=10$ Hz show a peaky distribution implying the phase synchrony between the neuronal populations; this may be compared with the noisy polar histograms in Fig.~\ref{fig:2}b.}
\label{fig:5}
\end{figure}

The inter-module connectivity parameters are set as in Table~\ref{tab:intermod_conn_params_ssvep} (see Section~\ref{sec:214}) when the cortical layers respond to periodic pulse input. Fig.~\ref{fig:3}a shows the time series of all three output populations responding to the 10 Hz periodic input. For a qualitative comparison, the outputs are overlaid pairwise. Note the membrane potentials of the populations is baseline corrected by subtracting their mean. The actual membrane potential values are TCR = -54.5, Py4 = -47.8 and Py6 = -35.3 mV. All outputs follow the input phase and frequency. The line plots of the PSD for $f=10$ Hz is shown in Fig.~\ref{fig:3}b, where the Py4 and Py6 populations now respond to the periodic input, unlike in the alpha rhythmic model state in Section~\ref{sec:31}. To demonstrate the suppression of alpha rhythm in this activated state, we have tested the model with just noise input i.e.\ no periodic flickering stimuli. The line plots of PSD in Fig.~\ref{fig:3}c show a $\frac{1}{f}$ (`inverse frequency') characteristic that is typical of a low amplitude noisy input; this is confirmed by an order of magnitude lower PSD amplitude compared to Fig.~\ref{fig:2}a. Fig.~\ref{fig:3}d show the Herrmann-like plots for model response to progressively increasing periodic input frequencies $f=1 - 30$ Hz and amplitudes $k=1 - 15$ mV. Both cortical populations Py4 and Py6 respond faithfully to the input frequencies. The maximum power within the fundamental frequency is most effective in the Py4 between 8 - 20 Hz, and in the Py6 between 5 - 15 Hz.
Fig.~\ref{fig:4}a shows a comparison of the synchronisation indices using PLV, NSE ($\rho_{norm}$) and CPI ($\lambda$) applied pairwise on the phase differences between the output populations. For input pulse amplitudes $k \gtrapprox 5$ mV, CPI and PLV approaches a maximum for all three pairwise combinations. This agrees with our qualitative inspection of the time series. In Fig.~\ref{fig:4}b the PLV values are used to plot the Arnold tongue formation showing the phase locked bounds on the amplitude-frequency plane. The phase slip plots in Fig.~\ref{fig:5}a are flatter for higher values of $k$ indicating increased phase locking between the pair. These plots are also aligned with the PLV measures indicating higher phase synchrony between the Py4 and Py6 than in the other pairs; the CPI index also shows similar synchronisation strength between the pairs. The polar histograms shown here for $f=10$ Hz and $k=5$ mV form tight clusters due to high phase locking corresponding to periodic input. The PLV comparison can also be observed in the Arnold tongue; note the colour bar indices. 
Overall, our model shows high phase synchronisation in the neuronal populations when stimulated with periodic pulse input. The fundamental frequencies in the model responses also follow the input frequencies. To confirm that this is indeed entrainment and not superposition, we provide jittered pulse as retinal input to the LGN. The results are presented below.

\begin{figure} [hb!]
\centering
\includegraphics[scale=0.95]{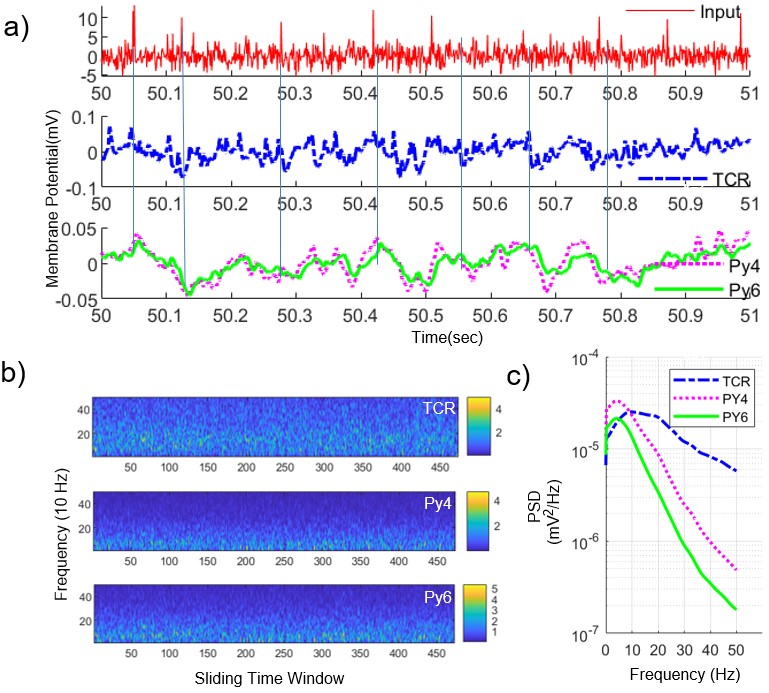}
\caption{(a) Time series outputs of the (middle) TCR and (bottom) Py4 and Py6 populations corresponding to (top) jittered flicker stimuli at 10 Hz provided as retinal input. Here, the mean membrane potential of the populations is baseline corrected for comparative scales. (b) The short-time-Fourier-Transform (STFT) show a lack of preference for any particular frequency throughout the simulation time. (c) Line power spectral density plots show a $1/f$ frequency characteristic for all three populations and no preference for the input frequency.}
\label{fig:6}
\end{figure}

\subsection{Response to Jittered Input}
\label{sec:33}

\begin{figure} [ht!]
\centering
\includegraphics[scale=1]{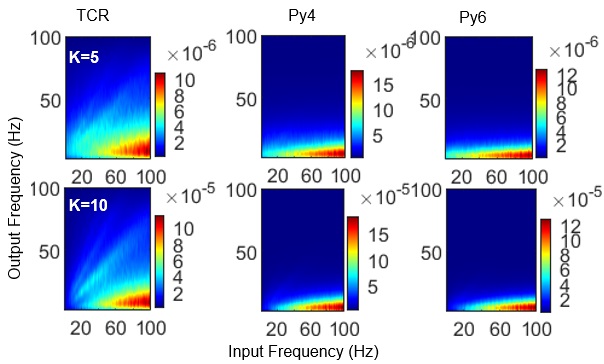}
\caption{Herrmann-like plots at two sampled pulse amplitudes of $k=5$ (top) and $k=10$ (bottom) are shown corresponding to jittered pulse as retinal input. The input frequencies are varied for a wider range $f=1 - 100$ Hz to confirm the lack of input frequency components in the cortical population Py4 and Py6 responses.}
\label{fig:7}
\end{figure}

The model is in the same state as that for periodic pulse input, and all inter-module parameters are as in Table~\ref{tab:intermod_conn_params_ssvep}. A qualitative inspection of the outputs in Fig.~\ref{fig:6}a shows that: first, the input frequency has a weak presence in the TCR output, and is absent in the Py4 and Py6 outputs; second, there is reduced phase alignment between the populations, and a lack of phase alignment between the retinal input and the TCR output, compared to those for the periodic input. Interested readers may note, the membrane potentials of the populations is baseline corrected by subtracting their mean. The actual membrane potential values are TCR = -54.5, Py4 = -47.8 and Py6 = -35.3 mV.
The line PSD plot in Fig.~\ref{fig:6}c shows peak power at frequencies that are much less than the input frequency of 10 Hz. STFT plots in Fig.~\ref{fig:6}b show noisy spectra for all the three populations. Herrmann-like plots for two different pulse amplitudes shown in Fig.~\ref{fig:7} further consolidate the absence of the jittered input frequency in the model response. This is in agreement with Notbohm \emph{et al}'s experimental observations.

\begin{figure}[hb!]
\begin{center}
\includegraphics[scale=0.6]{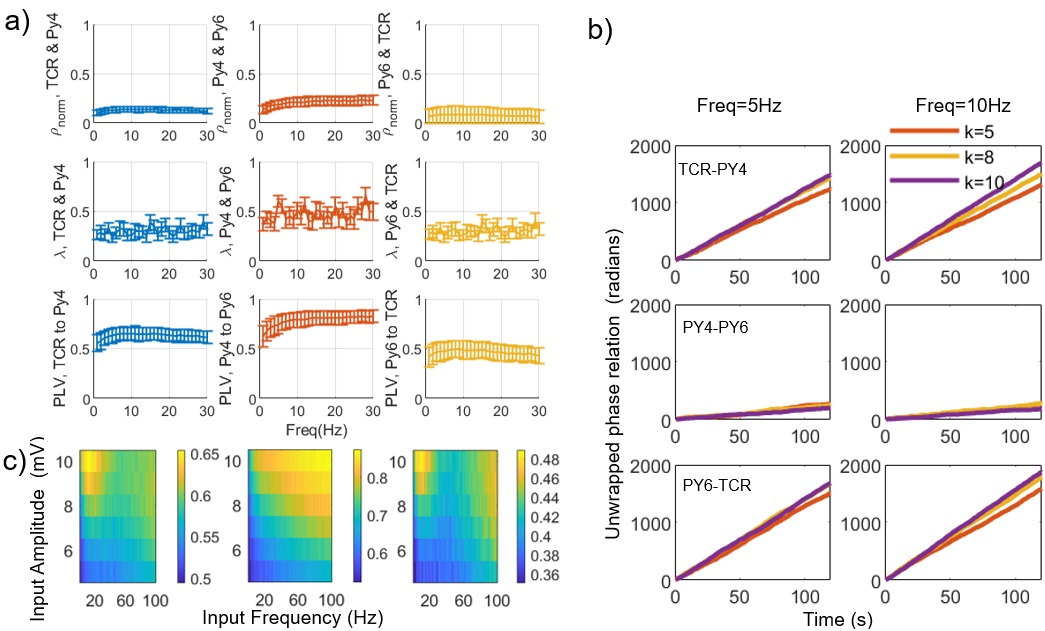}
\caption{(a) Synchronisation indices NSE ($\rho_{norm}$), CPI ($\lambda$) and PLV show lower mean values across all jittered flicker input frequencies compared to periodic flicker inputs. (b) Phase slip plots corresponding to increasing input frequencies and amplitudes show reduced phase synchrony among the populations compared to periodic flicker input. (c) The PLV heatmaps show a very narrow tongue formation for the left (TCR and Py4) and right (Py6 and TCR) around the area of the alpha peak frequencies.}
\label{fig:8}
\end{center}
\end{figure}

Compared to those of periodic inputs (see Fig.~\ref{fig:4} and Fig.~\ref{fig:5}), the phase synchronisation indices in Fig.~\ref{fig:8}a show lower values and the phase slip plots in Fig.~\ref{fig:8}b have steeper gradients even for high input amplitudes. The CPI indices are noisy and can be considered as a good indicator for detecting the jittered nature of the input. The Arnold tongue plots in Fig.~\ref{fig:8}c show the phase locking regions for higher frequency range for input amplitudes $k=5 - 10$. The TCR-Py4 and Py4-Py6 phase relations show weak synchronisation for higher amplitudes and in the region $\lessapprox 30$ Hz (compare the colour bar indices with Fig.~\ref{fig:4}b). However, the Py4-Py6 combination synchronisation indices with higher jitter frequency and amplitude are comparable to those for periodic input, albeit for lower frequencies. This is in spite of the time series and frequency plots showing little evidence of input phase and frequency in the input. This confirms the PLV synchronisation index is not sensitive to the noise in the data i.e.\ it includes noisy in-phase oscillations. In contrast the CPI seems to be more sensitive to noise, and shows high variance around the mean across the 20 simulation trials, the mean being significantly ($\gtrapprox 50\%$) reduced for all pairwise phase relations. The NSE plots, once again, convey little information about the synchronisation strengths.

Overall, our results show that corresponding to jittered pulse inputs, the frequency of the output populations do not follow those of the inputs, and there is reduced phase synchrony between the output populations. This result implies that the model response is not a superposition (i.e.\ linear combinations) of its inputs, but is affected by the nonlinearity in the model structure.

\begin{figure}[ht!]
\begin{center}
\includegraphics[scale=0.68]{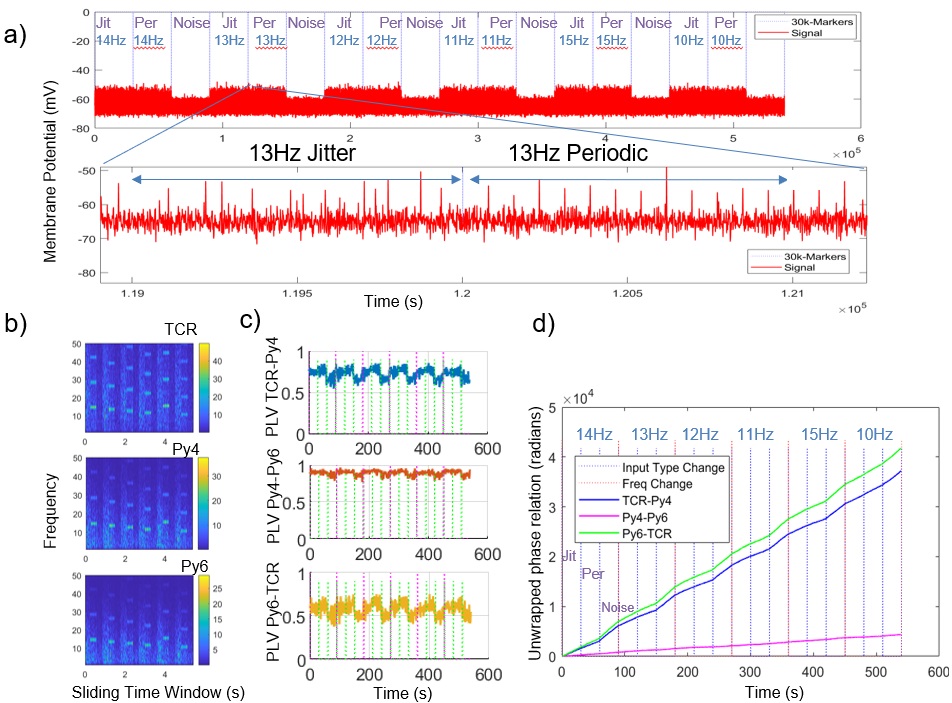}
\caption{(a) The mixed signal input time series generated for $540$ s. Each 90 s of the input signal is divided into 30 s of Jitter, Periodic and Noise conditions. The frequency for the Jitter and Periodic case is nonrepetitive and drawn randomly from a 10 to 15Hz bucket. 1 s on either side of 120 s slot is zoomed in to show the transition from Jitter to Periodic condition. (b) The STFT for TCR, Py4 and Py6 population responses to the mixed signal input. (c) The PLV synchronisation index measured for output population responses corresponding to the mixed signal input. All measures are computed over short-time windows of $1000$ ms with no overlap of adjacent windows. (d) Phase slip plots show phase synchrony between the thalamocortical populations with periodic input (plateaus) and disappearance of synchrony with jittered input and noise (ramps).}
\label{fig:9}
\end{center}
\end{figure}
\subsection{Response to Mixed Input}
\label{sec:34}

The mixed input signal is shown in Fig.~\ref{fig:9}a. Its STFT in Fig.~\ref{fig:9}(b--d) show three distinct patterns: the fundamental frequency and harmonics for periodic input that is similar to our observations in Section~\ref{sec:32}; a uniform spectrum with high average power for jittered input that is similar to our observations in Section~\ref{sec:33}; low output power compared to the responses to visual flickering stimuli (both periodic and jitter) corresponding to noise input.

To account for the continually changing nature of the input, all three phase synchronisation indices are computed over short time windows where the window size is 1 s with no overlap between adjacent windows. Overall nature of all measures in Fig.~\ref{fig:9}(c-d) are pulsating, where the trough corresponds to noise, the crest correspond to periodic input, and the ramp is due to a relatively higher phase synchrony for jittered input compared to noise. The crests in the PLV plots in Fig.~\ref{fig:9}c are more reflective of those in Fig.~\ref{fig:4}c, and confirm their robustness to noise in the input signal. The NSE and CPI metrics (not shown here) have much lesser values at the crest, compared to those for their pure periodic counterparts, implying slow recovery to varying synchronisation in the outputs. The phase slip plots in Fig.~\ref{fig:9}d have a pulsating nature, where the plateaus correspond to the periodic part of the signals and the ramps to the noise and jitter. We note here that the phase slip plots agree with the PLV synchronisation index

\begin{figure}[ht!]
\begin{center}
\includegraphics[scale=0.41]{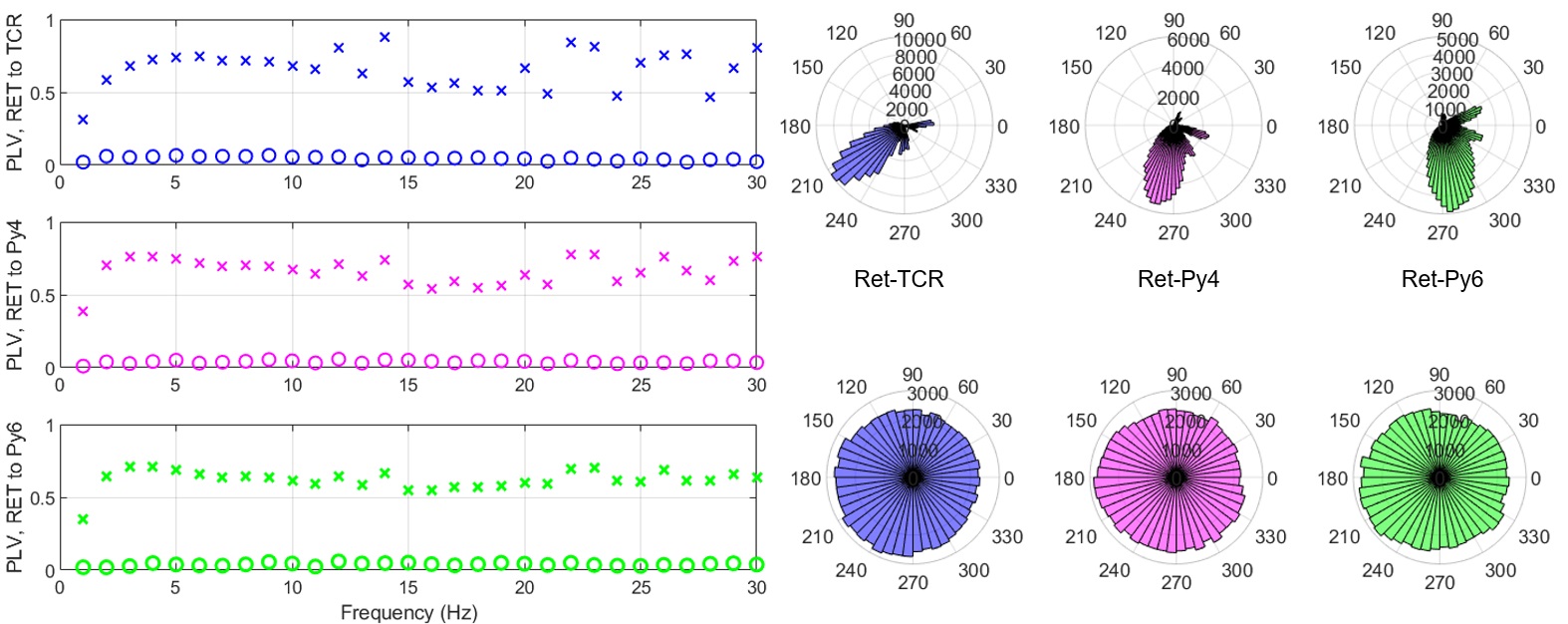}
\caption{(a) PLV and (b) polar histogram for relative phases between each network output and the periodic flicker input (marked as `x') versus jittered flicker input (marked as `o'). All populations are phase synchronised with periodic input. The phase synchronisation disappears with jittered input.}
\label{fig:10}
\end{center}
\end{figure}

Thus far, we have investigated phase synchronisation between neuronal populations in the network. In Fig.~\ref{fig:10}, we present the phase synchronisation between each network output and the flicker input. Comparison in synchronisation between the periodic vs jittered flicker inputs show much reduced phase synchrony with jittered input using PLV. The polar histograms also confirm this observation. We do not show the CPI and NSE measures as they are affected by the relatively high noise content in the input and do not provide reliable results.

In summary, our results show that SSVEP-like signals are produced in the cortical populations corresponding to periodic pulse retinal input and there is high phase synchrony between the network populations as well as with the input. In contrast, corresponding to jittered pulse inputs, the SSVEP-like signals disappear, all phase synchronisation indices are much reduced compared to those for periodic input. From this, we may infer that the phase synchrony corresponding to periodic input is due to entrainment of the population responses by the retinal input. Thus, our model is validated by the observations made in~\cite{notbohm2016modification}.
\section{Discussion}
\label{sec:4}
Instances of phase and frequency entrainment are ubiquitous in biology, for example the circadian rhythms are due to entrainment of sleep-wake cycles by `environmental cues' such as sunlight~\cite{ROSENBLUM2001279}. At the same time, entrainment of biological signals by extrinsic inputs are used to treat neurological disorders, for example both frequency and phase entrainment of neuronal responses occur during deep brain stimulation (DBS)~\cite{agnesi15} (DBS is brain stimulation therapy for symptomatic treatment of Parkinsonian tremor). Recent findings show that DBS can also be used to treat epilepsy~\cite{wang20}. Phase synchronisation in neuronal populations is also believed to play a `pivotal role' in memory processes in humans~\cite{fell11}. 

Here, our interest has been in quantifying phase synchronisation in a biologically inspired population neural network that is parameterised to simulate SSVEP-like signals. We have demonstrated phase entrainment of the \textit{in silico} model output by periodic flicker input using the following synchronisation indices: phase locking value (PLV), normalised Shannon entropy (NSE), and conditional probability index (CPI). The PLVs are further used to plot the Arnold tongue, which helps in visualising the range(s) of input signal frequencies and amplitudes that provide the maximum phase locking in the circuit. In addition, we have used polar histograms and phase slip plots for easy visualisation of the phase synchrony, or its absence, in the neuronal populations. The objective metrics and visualisation are similar to those in~\cite{notbohm2016modification} as well as other researches that have defined and demonstrated objectively the degree of phase synchronisation in biological signals~\cite{tass1998,ROSENBLUM2001279,lowet2016quantifying,Lachaux1999}.

Biologically inspired population neural networks are a widely accepted means to simulate and understand both healthy and abnormal brain states and functions \textit{in silico}.
One such biologically inspired population neural network is the lumped parameter model, which is also known as neural mass model (NMM)~\cite{david03}. The nomenclature is derived from the `neural mass'~\cite{freemanbook} concept, a group of tightly packed neurons at mesoscopic scale that may be considered as a point mass for all practical purposes.
Neural mass models are defined as a set of ordinary differential equations that are used to simulate and understand local field potentials (LFP) and scalp electroencephalogram (EEG) signals of neurological conditions such as epilepsy~\cite{piotr04,goodfellow14}, Parkinsonian tremor~\cite{marreiros2013}, or studying brain states such as consciousness~\cite{coalia2019}, as well as to understand brain stimulation~\cite{julien2017}. Also, these \textit{in silico} models are used to simulate functional Magnetic Resonance Imaging (fMRI)~\cite{friston00} and Blood Oxygen Level Dependant (BOLD) signals~\cite{soterobold07}. Neural field models, which are conceptually similar to neural mass models but have additional spatio-temporal characteristics using partial differential equations, are used to simulate and understand brain states such as anesthesia~\cite{axelhutt2011}, sleep-wake cycles~\cite{robinson2015}, as well as to make testable predictions on brain stimulation~\cite{julien2017}. We have been working with neural mass \textit{in silico} models to understand slowing of alpha rhythms as a definitive marker in the EEG of Alzheimer disease, as well as to understand the neuronal dynamics underpinning awake resting state alpha rhythms~\cite{Bhattacharya2011}. The \textit{in silico} model used in this study is an extension of our previous works.

The thalamocortical brain circuit subserving the occipital cortex are known to be vital in the generation of alpha rhythms, oscillations within the 8 -- 13 Hz frequency band that are observed at the O1 and O2 scalp electrodes of EEG recorded from subjects in an awake but resting state with eyes closed, for example before falling asleep. The alpha rhythms are known to disappear when the subjects open their eyes. The exact peak within the alpha band vary between individuals; the Individual Alpha Frequency (IAF) is an important measure in experimental studies that try to understand brain states and functions based on alpha rhythms. Alpha rhythms are one of the most intriguing oscillatory band that are prominent not only in a state of rest, but also recorded in attentive and sensory processing brain states~\cite{basar97,buzsaki2005}. Visual detection and performance is reported to be modulated by alpha rhythms~\cite{ergenoglu04}. Also, resonance within the alpha band and its harmonics is reported in SSVEP recordings~\cite{vialatte2010}. Besides its prevalence in different brain states of normal healthy adults, alpha rhythms form important biomarkers in the EEG of several neurological disorders; for example in Alzheimer disease and migraine, slowing of dominant frequency in the alpha range is a known phenomenon~\cite{vialatte2010}. SSVEPs are also known to be affected in several neurological disorders such as migraine, schizophrenia and epilepsy. The occipital cortex, from where both SSVEP and alpha rhythms are recorded, is the seat of the visual cortex. It is thus not surprising that \textit{in silico} models of the thalamocortical circuit in the visual pathway are used to understand neuronal dynamics underpinning both alpha rhythms~\cite{Bhattacharya2011} and SSVEP~\cite{Labecki2016,robinson2015}. The \textit{in silico} model in this work is informed by the thalamocortical circuit that form the visual information pathway of the brain and subserve the occipital cortex (see Section~\ref{sec:21}). It consists of three modules, viz.\ the thalamic module, the cortical Layer 4 module, and the cortical Layer 6 module. The output or the network are the responses of each module viz.\ the Thalamocortical Relay (TCR) population in the thalamic module, and the Pyramid populations of the Layer 4 (Py4) and Layer 6 (Py6) modules, which are analysed for frequency domain behaviour and phase synchrony.

Our results (see Section~\ref{sec:3}) indicate phase entrainment in our \textit{in silico} model in response to periodic flicker stimuli that mimic the ensemble membrane voltage of retinal spiking neurons. The phases as well as frequencies of all three populations are aligned with those of the periodic stimuli. In addition, the \textit{in silico} model allows insight into the phase synchronisation between the thalamocortical populations in the visual pathway corresponding to SSVEP-like model response. Both PLV and CPI metrics detect weak phase synchronisation in the presence of noise, and are more reliable in our work compared to NSE. For mixed signal inputs with increased noise, the CPI performed worse than PLV, but better than NSE. The phase synchronisation in all populations increase with higher amplitude of the retinal periodic pulse input. The response changes dramatically when the inter-pulse period of the retinal input is jittered; the phase synchronisation indices drop significantly and the output frequency no more follows that of the input. We note that if the model response were to be the superposition of the inputs, then, any variation of the input pulse frequencies would provide an additive response at the output. This is clearly not the case. Thus, the model response simulates the experimental observations in~\cite{notbohm2016modification} reporting SSVEP-like response as entrainment, rather than superposition, of the default brain oscillation by the input periodic stimuli. 

In addition, our \textit{in silico} study has indicated the role of inter-module synaptic connectivities in simulating resting brain state that is conducive to generating alpha rhythms, versus brain state corresponding to SSVEP-like signal generation. In a visual discrimination task between healthy subjects with IAF between 9-11 Hz, it was reported that high amplitude alpha rhythms correspond to good memory performance; conversely, good perceptual performers had low alpha rhythm amplitudes. Regardless of performance, all good performers, both memory and perceptual, had high phase locking within the alpha band. Furthermore, the study indicated that low alpha amplitudes, i.e. good perceptual performance, correspond to an active state of the cortex, while high alpha amplitudes correspond to a cortical deactivated state. We note that the overall feedforward and feedback connectivities between the three modules in our network had to be reduced to generate alpha rhythm. We consider this as a simulated awake resting state with eyes closed, when the cortex is deactivated with low inter-layer and inter-region connectivity and the retinal input is a random noise. In this state, the cortical layers do not respond to periodic retinal input even for high pulse amplitudes. Phase locking is observed only at the respective alpha peak of each neuron population, but falls off on both sides outside the peak frequency. Next, we increased all the inter-layer and inter-module connectivities that were earlier set to low values. Both cortical layers now responded to periodic input, even for low pulse amplitudes. We consider this as a state of cortical activation, when the model comes out of its resting state. Phase locking is high in all populations corresponding to periodic input. Such increase of phase synchronisations in neuronal populations with increased synaptic coupling strength is also reported in other \textit{in silico}  studies~\cite{breakspear03}. Thus, our \textit{in silico} study predicts that distinct brain states underpin resting state alpha rhythms and SSVEP-like signals. 

\section{Conclusion}
\label{sec:5}
The novelty of this work is two-fold: first, we have used a set of objective measures to quantify phase synchronisation in an \textit{in silico} model of the thalamocortical pathway in vision; second, we have validated a two-layer architecture of the visual cortex, interfaced with an existing LGN architecture, with experimental data demonstrating entrainment of neuronal dynamics that underpin SSVEP. 

Readers may note that the purpose of this work is not to make a thorough analysis of the \textit{in silico} model dynamics; rather, our goal is to simulate phase entrainment corresponding to SSVEP-like output. A detailed investigation of the parameter space using optimisation techniques will be applied to this model as a future work. Another brevity in this work is that m and n are always set to 1 for measuring phase synchrony. Thus, all our results are a study of $1:1$ phase locking in model responses. We will explore further combinations of $m:n$ in future work.

In conclusion, we note that entrainment by extrinsic stimuli such as in DBS is used to treat neurological conditions like Parkinsonian tremor and epilepsy, although the neuronal underpinnings are still a matter of research. This work demonstrates the usability of population neural networks to understand neuronal attributes that underpin brain entrainment. Based on our observations, we speculate that population neural networks validated by experimental data can make testable predictions for entrainment via brain stimulation as a therapeutic treatment in neurological conditions.

\section{Acknowledgements}
\label{sec:6}
This research is supported by the Birla Institute of Technology and Science (BITS) Pilani Goa Campus Grants BPGC/RIG/2018-19 and GOA/ACG/2019-20/Oct/02 to BSB. SS is supported by the BITS Pilani Goa Campus Institute Fellowship awarded towards her Doctoral Research.


 \bibliographystyle{elsarticle-num} 
 \bibliography{cas-refs}

\begin{thebibliography}{10}
\expandafter\ifx\csname url\endcsname\relax
  \def\url#1{\texttt{#1}}\fi
\expandafter\ifx\csname urlprefix\endcsname\relax\def\urlprefix{URL }\fi
\expandafter\ifx\csname href\endcsname\relax
  \def\href#1#2{#2} \def\path#1{#1}\fi

\bibitem{Michel2020}
M.~Wälti~J, D.~G, N.~Wenderoth, {Assessing Rhythmic Visual Entrainment and
  Reinstatement of Brain Oscillations to Modulate Memory Performance},
  Frontiers in Behav. Neuroscience (2020).

\bibitem{wang20}
W.~Zhaoxiang, F.~Zhouyan, Y.~Yue, Z.~Lvpiao, Suppressing synchronous firing of
  epileptiform activity by high-frequency stimulation of afferent fibers in rat
  hippocampus, CNS Neurosci Ther. 27 (2020) 352--362.

\bibitem{vialatte2010}
F.~B. Vialatte, M.~Maurice, J.~Dauwels, A.~Cichocki, {Steady-state visually
  evoked potentials: focus on essential paradigms and future perspectives},
  Progress in Neurobiology 90~(4) (2010) 418--438.

\bibitem{norcia2015steady}
A.~M. Norcia, L.~G. Appelbaum, J.~M. Ales, B.~R. Cottereau, B.~Rossion, The
  steady-state visual evoked potential in vision research: A review, Journal of
  Vision 15~(6) (2015) 4--4.

\bibitem{notbohm2016modification}
A.~Notbohm, J.~Kurths, C.~S. Herrmann, Modification of brain oscillations via
  rhythmic light stimulation provides evidence for entrainment but not for
  superposition of event-related responses, Frontiers in Human Neuroscience 10
  (2016) 10.

\bibitem{basar97}
E.~Basar, M.~Schurmann, C.~Basar-Eroglu, S.~Karakas, Alpha oscillations in
  brain functioning: an integrative theory, International Journal of
  Psychophysiology 26 (1997) 5--29.

\bibitem{niedermeyer97}
E.~Niedermeyer, Alpha rhythms as physiological and abnormal phenomena,
  International Journal of Psychophysiology 26 (1997) 31--49.

\bibitem{piotr04}
P.~Suffczy\'nski, S.~Kalitzin, F.~L. da~Silva, Dynamics of non-convulsive
  epileptic phenomena modelled by a bistable neuronal network, Neuroscience 126
  (2004) 467--484.

\bibitem{david03}
O.~David, K.~J. Friston, A neural mass model for {MEG}/{EEG}: coupling and
  neuronal dynamics, NeuroImage 20 (2003) 1743--1755.

\bibitem{Labecki2016}
M.~Labecki, R.~Kus, A.~Brzozowska, T.~Stacewicz, B.~Sen~Bhattacharya,
  P.~Suffczynski, Nonlinear origin of ssvep spectra — a combined experimental
  and modeling study, Front. Comput. Neurosci. 10 (2016) 1--10.

\bibitem{Bhattacharya2011}
B.~Sen~Bhattacharya, D.~Coyle, L.~Maguire, {A thalamo–cortico–thalamic
  neural mass model to study alpha rhythms in Alzheimer’s disease}, Neural
  Networks 24 (2011) 631--645.

\bibitem{frontiers2013}
B.~Sen~Bhattacharya, Implementing the cellular mechanisms of synaptic
  transmission in a neural mass model of the thalamo-cortical circuitry,
  Frontiers in Computational Neuroscience 7 (2013) 81.

\bibitem{bhattacharya2016causal}
B.~Sen~Bhattacharya, T.~P. Bond, L.~O'hare, D.~Turner, S.~J. Durrant, Causal
  role of thalamic interneurons in brain state transitions: a study using a
  neural mass model implementing synaptic kinetics, Frontiers in Computational
  Neuroscience 10 (2016) 115.

\bibitem{mahajan2021}
P.~Mahajan, A.~Rane, S.~Sasi, B.~Sen~Bhattacharya, {Quantifying Synchronization
  in a Biologically Inspired Neural Network}, In Proceedings: International
  Joint Conference on Neural Networks (IJCNN) (2021) 1--8.

\bibitem{ROSENBLUM2001279}
M.~Rosenblum, A.~Pikovsky, J.~Kurths, C.~Schäfer, P.~Tass, Phase
  synchronization: From theory to data analysis, in: F.~Moss, S.~Gielen (Eds.),
  Neuro-Informatics and Neural Modelling, Vol.~4 of Handbook of Biological
  Physics, North-Holland, 2001, Ch.~9, pp. 279 -- 321.

\bibitem{tass1998}
P.~Tass, M.~G. Rosenblum, J.~Weule, J.~Kurths, A.~Pikovsky, J.~Volkmann,
  A.~Schnitzler, H.-J. Freund, Detection of $\mathit{n}:\mathit{m}$ phase
  locking from noisy data: Application to magnetoencephalography, Phys. Rev.
  Lett. 81 (1998) 3291--3294.

\bibitem{Lachaux1999}
J.-P. Lachaux, E.~Rodriguez, J.~Martinerie, F.~J. Varela, Measuring phase
  synchrony in brain signals, Human Brain Mapping 8~(4) (1999) 194--208.

\bibitem{lowet2016quantifying}
E.~Lowet, M.~J. Roberts, P.~Bonizzi, J.~Karel, P.~De~Weerd, Quantifying neural
  oscillatory synchronization: a comparison between spectral coherence and
  phase-locking value approaches, PloS One 11~(1) (2016) e0146443.

\bibitem{shermanbook}
S.~M. Sherman, R.~W. Guillery, Exploring the thalamus, 1st Edition, Academic
  Press, New York, 2001.

\bibitem{sherman2001-ciruit}
S.~Sherman, {Sherman SM. Tonic and burst firing: dual modes of thalamocortical
  relay}, Trends Neurosci. 24~(2) (2001) 122--26.

\bibitem{hanslmayr2005}
S.~Hanslmayr, W.~Klimesch, P.~Sauseng, W.~Gruber, M.~Doppelmayr,
  R.~Freunberger, T.~Pecherstorfer, {Visual discrimination performance is
  related to decreased alpha amplitude but increased phase locking},
  Neuroscience Letters 275 (2005) 64--68.

\bibitem{vindiola2014}
M.~Vindiola, J.~Vettel, S.~Gordon, P.~Franaszczuk, K.~McDowell, Applying eeg
  phase synchronization measures to non-linearly coupled neural mass models.,
  Journal of neuroscience methods 226 (01 2014).

\bibitem{sherman2013functional}
S.~M. Sherman, R.~W. Guillery, Functional connections of cortical areas: a new
  view from the thalamus, MIT Press, 2013.

\bibitem{douglas2007mapping}
R.~J. Douglas, K.~A. Martin, Mapping the matrix: the ways of neocortex, Neuron
  56~(2) (2007) 226--238.

\bibitem{binzegger2004quantitative}
T.~Binzegger, R.~J. Douglas, K.~A. Martin, A quantitative map of the circuit of
  cat primary visual cortex, Journal of Neuroscience 24~(39) (2004) 8441--8453.

\bibitem{pritchard92}
W.~S. Pritchard, D.~W. Duke, Measuring chaos in the brain: A tutorial review of
  nonlinear dynamical eeg analysis, International Journal of Neuroscience 67
  (1992) 31--80.

\bibitem{buzsaki2005}
G.~Buzs\'aki, Rhythms of the Brain, Oxford University Press, 2004.

\bibitem{glass2001synchronization}
L.~Glass, Synchronization and rhythmic processes in physiology, Nature
  410~(6825) (2001) 277--284.

\bibitem{Herrmann2001}
C.~S. Herrmann, Human eeg responses to 1-100 hz flicker: resonance phenomena in
  visual cortex and their potential correlation to cognitive phenomena, Exp
  Brain Res 137 (2001) 346--53.

\bibitem{robinson2015}
P.~A. e.~a. Robinson, A Multiscale “Working Brain” Model, Vol.~14,
  Springer,Cham, 2015, Ch.~5, pp. 107--140.

\bibitem{agnesi15}
F.~Agnesi, A.~Muralidharan, K.~Baker, J.~Vitek, M.~Johnson, Fidelity of
  frequency and phase entrainment of circuit-level spike activity during dbs, J
  Neurophysiol 114 (2015) 825--34.

\bibitem{fell11}
J.~Fell, N.~Axmacher, The role of phase synchronization in memory processes,
  Nat Rev Neurosci. 12 (2011) 105--18.

\bibitem{freemanbook}
W.~J. Freeman, Mass action in the nervous system, 1st Edition, Academic Press,
  New York, 1975.

\bibitem{goodfellow14}
Y.~Wang, M.~Goodfellow, P.~Taylor, G.~Baier, Dynamic mechanisms of neocortical
  focal seizure onset, PLOS Computational Biology 10 (2014) e1003.

\bibitem{marreiros2013}
A.~Marreiros, H.~Cagnan, R.~Moran, K.~Friston, P.~Brown, Basal ganglia-cortical
  interactions in parkinsonian patients, Neuroimage 66 (2013) 301--10.

\bibitem{coalia2019}
S.~Bensaid, J.~Modolo, I.~Merlet, F.~Wendling, P.~Benquet, Coalia: A
  computational model of human eeg for consciousness research, Frontiers in
  Systems Neuroscience 13 (2019) 59.

\bibitem{julien2017}
M.~Faten, M.~Julien, R.~Fanny, D.~Gabriel, B.~Arnaud, B.~Pascal, W.~Fabrice,
  Model-guided control of hippocampal discharges by local direct current
  stimulation, Scientific Reports 7 (2017).

\bibitem{friston00}
K.~Friston, A.~Mechelli, R.~Turner, C.~Price, Nonlinear responses in fmri: The
  balloon model, volterra kernels, and other hemodynamics, NeuroImage 12~(4)
  (2000) 466--477.

\bibitem{soterobold07}
C.~S. Roberto, T.-B. Nelson, Modelling the role of excitatory and inhibitory
  neuronal activity in the generation of the bold signal, NeuroImage 35~(1)
  (2007) 149–65.

\bibitem{axelhutt2011}
A.~Hutt (Ed.), Sleep and Anesthesia, Springer-Verlag New York, 2011.

\bibitem{ergenoglu04}
T.~Ergenoglu, T.~Demiralp, Z.~Bayraktaroglu, M.~Ergen, H.~Beydagi, Y.~Uresin,
  Alpha rhythm of the eeg modulates visual detection performance in humans,
  Brain Res Cogn Brain Res 20~(3) (2004) 376–86.

\bibitem{breakspear03}
M.~Breakspear, J.~Terry, K.~Friston, Modulation of excitatory synaptic coupling
  facilitates synchronization and complex dynamics in a nonlinear model of
  neuronal dynamics, Neurocomputing 52 (2003) 151--158.

\end{thebibliography}





\end{document}